\documentstyle[epsfig,twocolumn,aps,prl,epsf,floats]{revtex}
\begin{document}
\draft
\flushbottom
\twocolumn[
\hsize\textwidth\columnwidth\hsize\csname @twocolumnfalse\endcsname

\title{Tunneling density of states of high $T_c$ superconductors\\
  d-wave BCS model vs. $SU(2)$ slave boson model}

\author{Walter Rantner and Xiao-Gang Wen}

\address{Dept.\ of Physics, Massachusetts Institute of Technology,
Cambridge, Massachusetts 02139}
\widetext
\date{ 19 May\ 2000}
\maketitle
\tightenlines
\widetext
\advance\leftskip by 57pt
\advance\rightskip by 57pt

\begin{abstract}

Motivated by recent experimental measurements of the tunneling
characteristics of high T$_{c}$ materials using scanning tunneling
spectroscopy, we have calculated the IV and differential conductance curves
in the superconducting state at zero temperature. Comparing the two results
obtained via BCS-like d-wave pairing and the $SU(2)$ slave boson approach, 
we find that the slave-boson model can explain the asymmetric background
observed in experiments. The slave-boson model also predicts that
the height of the
conductance peak relative to the background
is proportional to the hole doping concentration $x$, at least for under-doped
samples. 
We also observe the absence of the van Hove singularity, and comment on
possible implications.
\newline
\newline
\end{abstract}
]
\tightenlines
\narrowtext
\section{Introduction}

Tunneling spectroscopy has been one of the fundamental tools in studying the
superconducting state of the high T$_{c}$ materials. In recent years it has
been possible to use the scanning tunneling microscope (STM) to perform
reproducible experiments on single crystal
Bi$_{2}$Sr$_{2}$CaCu$_{2}$O$_{8+\delta}$ cleaved in ultra high vacuum
\cite{Renner,Pan}.
In contrast to photoemission experiments, which are local probes in
wave-vector space, the STM is local in real space. Thus it 
does not provide any information that depends on momentum, however, it has much
higher energy resolution.  The density of states (DOS)
obtained from the $\frac{dI}{dV}$ curve is a direct fingerprint of
the single particle microscopic physics in the material and hence of some
importance in constraining the possible theoretical models explaining the
elusive high T$_{c}$ physics.

The STM spectra for the superconducting phase exhibit unusual structure in
the DOS when viewed in the light of BCS even if effects such as energy
dependence of the normal state density of states of sample and/or tip,
existence of bandwidth cutoffs, unequal work functions of tip and sample and
energy-dependent transmission probabilities are included \cite{Hirsch}.
One notable feature in the tunneling spectra 
is the asymmetric background with an enhancement for hole-tunneling into the
sample. This feature is reproduced very well in our calculation based on the
slave-boson theory. The slave-boson model also predicts that the
strength of the background in the tunneling spectra does not scale with
the doping $x$ while the sharp conductance peak scales linearly with $x$.
Thus by measuring the relative strength of the background and the sharp 
conductance peak as a function of doping $x$, 
one can distinguish the BCS theory and the slave boson
theory experimentally.

We also observe the absence of the van Hove singularity, and comment on 
possible implications.
One possible implication is particularly intriguing and consistent with
Photoemission results. That is the quasiparticles have a long lifetime,
$\tau > 0.5$/meV, below the superconducting gap, and a very short lifetime,
$\tau < 0.05$/meV, (spin-charge separation) above the superconducting
gap.

\section{d-wave BCS}

The differential conductance $\frac{dI}{dV}$ displays  in the simplest case
of constant DOS in the tip and energy independent transition probability the
single electron DOS in the sample. This reflects the ability of the material
to accommodate an extra electron or hole depending on the sample bias.
Within the tunneling Hamiltonian formalism the tunneling current is
given by \cite{Mahan}
\begin{eqnarray}\label{j}
  j_{T}  =  4\pi e \Gamma^2 \sum_{k,p}\int & d\omega &[
A_{L-}(\omega,p)A_{R+}(\omega + V,k) \nonumber\\
 & - & A_{L+}(\omega,p)A_{R-}(\omega + V,k)]
\end{eqnarray}
where the $A_{L}$s and $A_{R}$s are the spectral functions for the single
electron Green's functions in the tip (L) and sample (R) 
respectively. $V$ denotes the bias of the sample with
respect to the tip and $\Gamma$ is the tunneling matrix element assumed
independent of energy. Notice that positive $V$ corresponds to $e^{-}$
tunneling into the sample.

For a free fermion system which we suppose to represent the tip material we
have the standard form for the spectral function at zero temperature
$A_{L+}(\omega,p) = \Theta(\omega)\delta(\omega - \xi_{p})$
and $ A_{L-}(\omega,p) = \Theta(-\omega)\delta(\omega - \xi_{p})$,
were $\xi_{p} = \epsilon_{p} - E_{F}$ denotes the particle spectrum in the
tip with $E_{F}$ the Fermi energy. $\Theta(\omega)$ is the Heaviside 
step function where $\omega$ is measured with respect to the Fermi energy.

For the sample, we first consider the following single particle spectral
distribution at zero temperature.
\begin{eqnarray}\label{A+-dwave}
A_{R+}(\omega,p) &=& \Theta(\omega)u^2(p)\delta(\omega - E_{p}) \nonumber\\
A_{R-}(\omega,p) &=& \Theta(-\omega)v^2(p)\delta(\omega + E_{p})
\end{eqnarray}
 where $u^2(p) = \frac{1}{2}(1+\frac{\xi_{p}}{E_{p}})$ and $v^2(p) =
\frac{1}{2}(1-\frac{\xi_{p}}{E_{p}})$ are the BCS coherence factors $E_{p} =
\sqrt{\xi_{p}^2 + \Delta_{p}^2}$ is the dispersion relation of the
quasiparticles in the superconducting state and $\xi_{p} = \epsilon_{p} -
E_{F}$ with $\epsilon_{p}$ the dispersion relation in the normal state. Here
$\Delta_{p}$ denotes the gap function which is taken to have a d-wave
symmetry in reciprocal space.
Within the above approximation to the spectral functions we obtain the
following expressions for the single particle tunneling current 
\begin{eqnarray}\label{jdwave}
j_{T}|_{V>0} &=& 4e\pi\Gamma^2 N(E_{F})\sum_{k,E(k)\leq V}u^2(k)
\nonumber\\
j_{T}|_{V<0} &=& -4e\pi\Gamma^2 N(E_{F})\sum_{k,E(k)\leq
|V|}v^2(k)\nonumber
\end{eqnarray}

\begin{figure}[tb]
\epsfxsize=65mm
\centerline{ \epsfbox{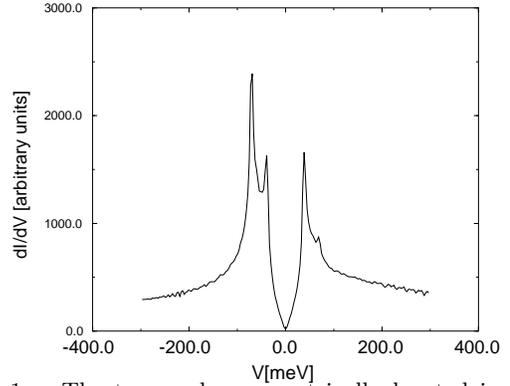} }
\caption{ The two peaks symmetrically located in height and energy with
respect to zero bias are the usual peaks arising from the gap structure in
the superconducting density of states. The two outer peaks are the remnants
of the van Hove singularity. The wiggles in the background are due to the
discreteness of k-space when performing numerical calculation. The discreteness
gets amplified by the derivative taken in obtaining $\frac{dI}{dV}$ from the
tunneling current (\ref{jdwave}). The resolution in voltage is 3meV}
\label{didv-dwave}
\end{figure}

In Fig. (\ref{didv-dwave}) we have plotted the resulting differential
conductance curve.  With
\begin{eqnarray}
\xi_{k} = t_{0} &+& t_{1}(\cos k_{x} + \cos k_{y}) +
t_{2}(\cos k_{x}\cos k_{y})\nonumber \\
&+& t_{3}(\cos 2k_{x}+\cos 2k_{y} + t_{4}(\cos 2k_{x}\cos k_{y}\nonumber \\
&+& \cos 2k_{y}\cos k_{x}) + t_{5}\cos 2k_{x}\cos 2k_{y} - E_{F}\nonumber
\end{eqnarray}
 as the dispersion relation  for the quasiparticles in the ab - plane
of the sample and the matrix elements chosen as follows
 $[t_{0},...,t_{5}]$~=~[0.1305,-0.2976,0.1636,-0.026,-0.0559,0.051] in eV,
this is a tight binding fit to angle resolved photo-emission spectroscopy
(ARPES) measurements performed by Norman {\it et al} \cite{Norman}. $E_{F}$ has
been adjusted to yield 10\% hole doping and $\Delta_{k} =
\Delta_{0}(\cos k_{x} - \cos k_{y})$ ( with $\Delta_{0} = 22meV$) has the d-wave
k-space symmetry mentioned previously.

With the hole doping at 10\%, the van Hove singularity, present in the
band structure, ends up on the hole side of the DOS very close to the Fermi
energy.
The fact that the van Hove singularity is close to the Fermi surface and
hence should show up in the low energy single particle physics can be seen
nicely in Fig. (\ref{didv-dwave}) in the form of the double peak
structure. The coherence factors which mix particle and hole density of
states lead to the van Hove singularity also showing up on the particle side
of the $\frac{dI}{dV}$ curve albeit with much smaller amplitude. 

\section{Slave bosons}

Next we consider the tunneling problem in the light of the $SU(2)$ slave boson
theory of Wen and Lee \cite{SU2meanfield}. It is commonly believed that the
simplest model that incorporates the strong correlation physics relevant
for the high $T_{c}$ cuprates is the t-J model. Due to the strong on-site
Coulomb repulsion energy the doubly occupied states should not contribute to
the low energy effective theory. Within the $SU(2)$ approach this constraint
is implemented via the introduction of a slave boson doublet.
The physical electron operator can then be written as an $SU(2)$ singlet.
Within this representation, the mean-field electron propagator is given by
the product of the boson and the fermion propagators and was calculated in
\cite{fourauthor}.

For our purpose we only need the $T=0$ spectral functions which can be read
off from the expression for the Green's function as
\begin{displaymath}\label{ A+SU(2)}
A_{R+}(\omega,k) = \Theta(\omega)[\frac{x}{2}u_{f}^2(k)\delta(\omega -
E^{f}(k))]
\end{displaymath}
for the particle part
\begin{eqnarray}\label{A-SU(2)}
A_{R-}(\omega,k) &=& \Theta(-\omega)\Big\{\frac{x}{2}v_{f}^2(k)\delta(\omega
+ E^{f}(k)) \nonumber \\
		+ \frac{1}{2N}\sum_{q}[u_{b}(q &-& k)u_{f}(q) +
v_{b}(q-k)v_{f}(q)]^2\nonumber \\
&\times &\delta(\omega + E^{f}(q) + E^{b}_{-}(q-k))  \\
		+ \frac{1}{2N}\sum_{q}[u_{b}(q &-& k)v_{f}(q) -
v_{b}(q-k)u_{f}(q)]^2\nonumber \\
&\times &\delta(\omega + E^{f}(q) + E^{b}_{+}(q-k))\Big\} \nonumber
\end{eqnarray}
for the hole part of the spectrum.
Here $N$ denotes the number of sites and $x$ is the hole doping concentration.
The remaining variables are defined as follows:
$u_{f,b}(k) =
\frac{1}{\sqrt{2}}\sqrt{1+\frac{\epsilon(k)^{f,b}}{|E^{f,b}|}}$,
$ v_{f,b}(k) =
\frac{1}{\sqrt{2}}\frac{\Delta^{f,b}(k)}{|\Delta^{f,b}(k)|}
\sqrt{1-\frac{\epsilon(k)^{f,b}}{|E^{f,b}|}} $,
$ E^{f}(k) = \sqrt{(\epsilon^{f}(k))^2 + (\Delta^{f}(k))^2}$ and 
$ E^{b}_{\pm}(k) = \pm\sqrt{(\epsilon^{b}(k))^2 + (\Delta^{b}(k))^2} -
\mu_{b}$, 
where $\epsilon^{f}(k),\Delta^{f}(k)$ are the fermion dispersion and gap
function respectively $\epsilon^{b}(k),\Delta^{b}(k)$ are the dispersions of
the bosons and $\mu_{b}$ is the boson chemical potential.

Notice that besides a coherent part for the spectral functions which
resembles the form of the BCS spectral weight (\ref{A+-dwave}) albeit scaled
by a factor of $\frac{x}{2}$ there is also an added incoherent contribution
to the hole part of the spectral function (\ref{A-SU(2)}).

To calculate the tunneling current we have used the fit to ARPES
measurements as dispersion for the fermions (spinons) and a nearest neighbor
tight binding dispersion
\begin{displaymath}
\epsilon^{b} = -2t^{b}(\cos k_{x}+\cos k_{y})
\end{displaymath}
for the bosonic degrees of freedom (holons) with $t^{b}$ the hopping matrix
element for the holons. It is important that we match the fermionic
band structure with the ARPES measurements since the fermions have a bigger
band mass and hence determine the dispersion relation seen in ARPES
\cite{SU2meanfield,Laughlin}  
The electrons measured in those experiments are thought of
(within spin-charge separating models) as bound states of the \emph{heavy}
spin degrees of freedom and the \emph{light} charge degrees of freedom.
Since we are interested in the low energy effective theory the details of
the broad dispersion for the holons (charge degrees of freedom) are not
crucial and hence have been chosen as simple as possible.
Furthermore to arrive at equations (\ref{A-SU(2)}) we assumed boson
condensation of the holons.

With the above expressions for the spectral functions in the sample we can
calculate the tunneling current using equation (\ref{j}) as
\begin{eqnarray}\label{jSU2}
j_{T}|_{V>0} &=& 4e\pi\Gamma^2 N(E_{F})\sum_{k,E(k)\leq V}
\frac{x}{2}u_f^2(k) \nonumber\\
j_{T}|_{V <0} &=& -4e\pi\Gamma^2 N(E_{F})\sum_{k,E(k)\leq |V|}
\frac{x}{2}v_f^2(k) \\ \nonumber \\
 &+& \frac{1}{2N}\sum_{k,q}^{-}[u_{b}(q - k)u_{f}(q) +
v_{b}(q-k)v_{f}(q)]^2\nonumber \\
		&+& \frac{1}{2N}\sum_{k,q}^{+}[u_{b}(q - k)v_{f}(q) -
v_{b}(q-k)u_{f}(q)]^2\nonumber
\end{eqnarray}
where 
$
\sum_{k,q}^{\pm}  =  \sum_{k,q}[\Theta(E^{f}(q) + E^{b}_{\pm}(q-k))
		  -    \Theta(E^{f}(q) + E^{b}_{\pm}(q-k) - |V|)]$.

\begin{figure}[tb]
\epsfxsize=4.5cm
\hskip 4.1cm \epsfbox{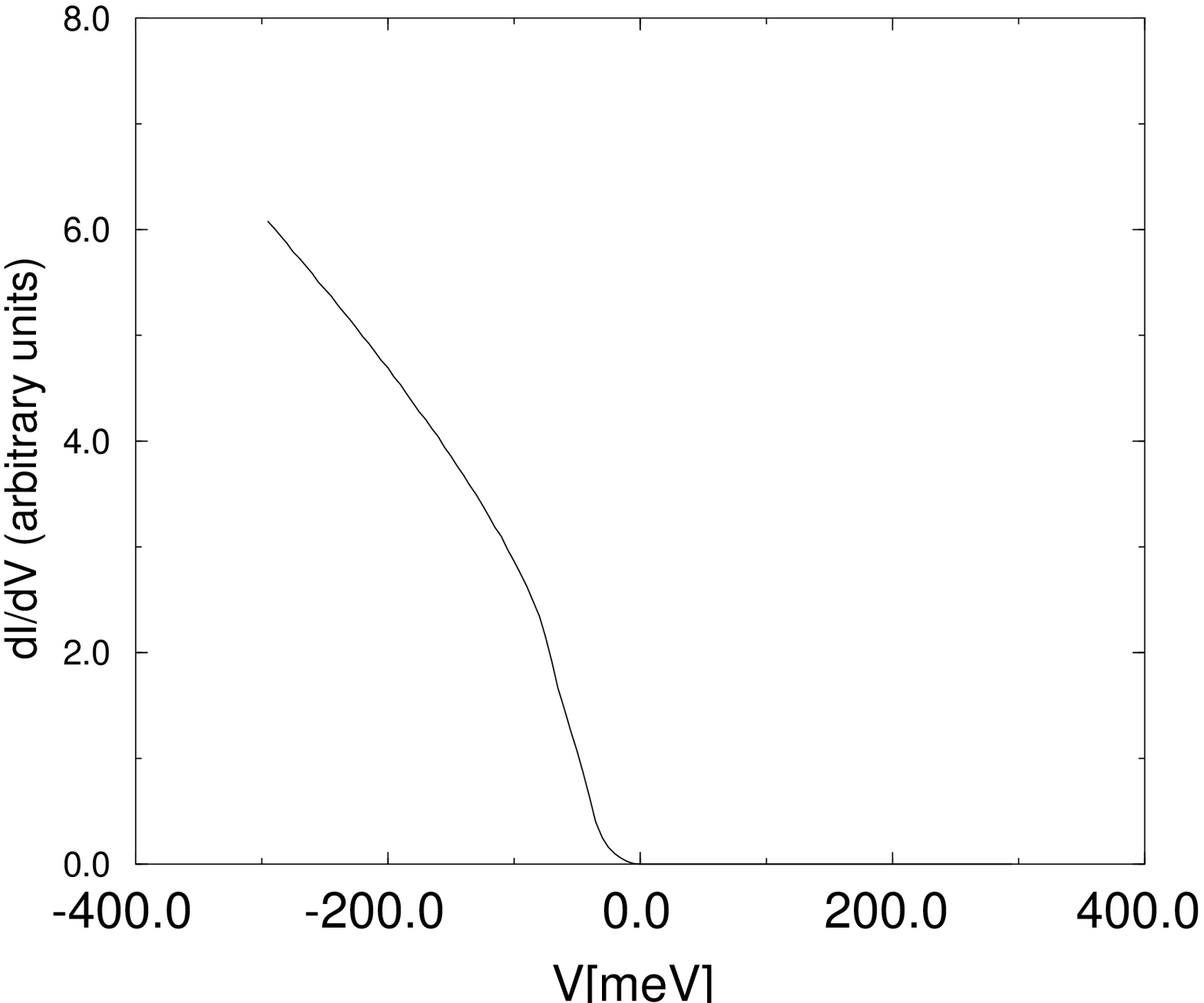} \\
\epsfxsize=3.5in
\vskip -4cm
\centerline{\epsfbox{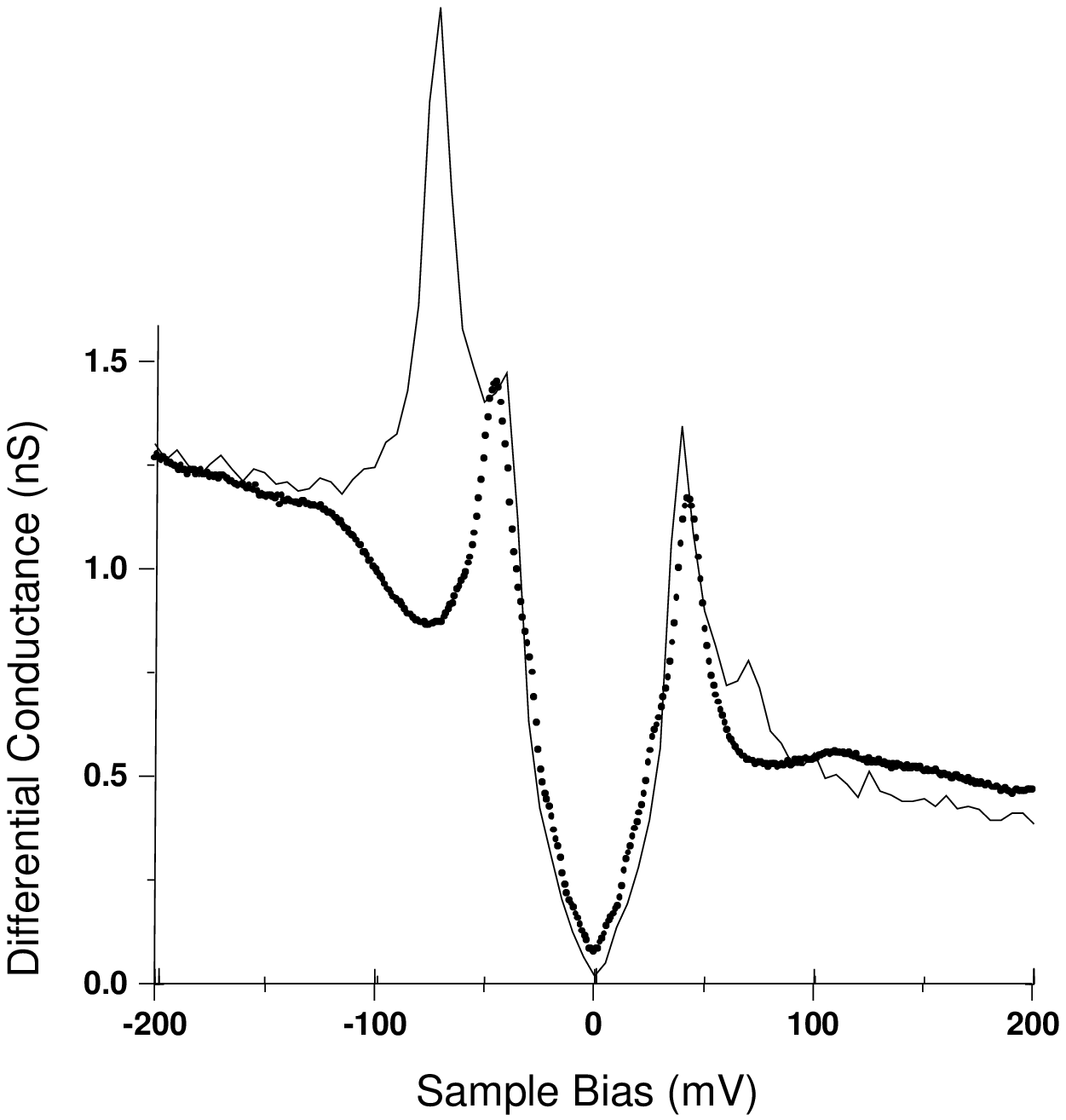} }
\caption{Here we have set $t_{1}^{f}/2t^{b}=1/2$.
$\Delta^{f} = \Delta_{0}^{f}(\cos k_{x}-\cos k_{y})$ and $\Delta^{b} =
\Delta_{0}^{b}(\cos k_{x}-\cos k_{y})$ with $\Delta_{0}^{f}/\Delta_{0}^{b}=1/2$
where $t_{1}^{f}~=~-297.6$~meV from the ARPES fit and $\Delta_{0}^{f} =
22$~meV. Notice that above ratios correspond to  $J/t = 1/2$ within the t-J
model. The solid curve is our slave-boson result. 
The wiggles in the background are more pronounced here as compared
with Fig. (\ref{didv-dwave}) due to smaller resolution in k-space in
calculating the convolution integrals (\ref{jSU2}). Here the resolution in
$V$ is 5 meV. The dots are the experimental result from Ref. [2].
The inset shows the incoherent part of the hole tunneling spectral weight. 
Notice that
the exact shape of this curve should not be taken too literally - see
discussion on boson band structure in the main text.
}
\label{didvSU2}
\end{figure}

When comparing
the two $\frac{dI}{dV}$ curves Fig. (\ref{didv-dwave}) and
(\ref{didvSU2}) one can see how the lowest energy physics is virtually 
identical. However on energy scales
bigger than $4\Delta_{0}$ a marked asymmetry in the background of the $SU(2)$
model $\frac{dI}{dV}$ shows up with an increase in the hole tunneling
spectral weight.

The inset of Fig. \ref{didvSU2} depicts the 
incoherent contribution to the hole tunneling
spectrum separately. Scaling the height of Fig. \ref{didv-dwave} by
$\frac{x}{2}$ and adding the incoherent hole contribution results in 
Fig. \ref{didvSU2}.
The increase in the hole tunneling spectral weight arises due to the fact
that removing an electron from the sample requires the recombination of the
spin and charge degrees of freedom into a single entity. This yields a
mixing in of the higher energy holon dispersion whose detailed form is not
known within the effective low energy theory.

Another feature so far not discussed is the
scaling with the hole doping of the conductance peak corresponding to
electrons tunneling into the sample. Comparing equations (\ref{jdwave}) and
(\ref{jSU2}) we see that within the $SU(2)$ model,
the peak height scales linearly with $x$
whereas there is no dependence of the peak height on doping within the d-wave
approach.
The doping dependence within $SU(2)$ arises from the reduction of the overlap
of the electron in the tip with the quasiparticle (as a bound state of holons
and spinons) in the sample which crudely speaking means that an electron can
only enter the sample on empty sites and then 'decay' into its constituent
parts.
The only doping dependence within the d-wave approach arises due to the
chemical potential which dictates the separation of the double peak
structure but not its height.
The linear scaling with $x$ within the $SU(2)$ slave boson mean-field theory
discussed here should be taken more as a qualitative than exact
quantitative prediction, since it is a mean-field result.
Recent photoemission experiments\cite{Ding} by Ding {\it et al} observed a linear $x$ dependence of the quasiparticle
peak, which fits the mean-field result of the $SU(2)$ theory very well.

Thus, at the mean-field level we have found qualitative different behaviors with
regards to the $x$ dependence of the hole tunneling background and the
electron tunneling peak within $d$-wave BCS and the $SU(2)$ slave boson
theory.  It is this difference in doping dependence, which should be
experimentally testable and hence yield to a feature distinguishing between
the two models. 

Furthermore notice that the DOS contains singularities (for both BCS
and the slave boson model) at the electron tunneling peak. The curvature
of the measured $dI/dV$ curve at these peaks should give us an upper bound on 
the quasiparticle decay rate at the energy scale of the superconducting gap.
Based on new experimental data by Pan {\it et al}\cite{Pan}, 
the  quasiparticle decay rate 
can be as small as a few $meV$ even for quasiparticles with energy as high as
$40meV$. This is very different from the normal state where the quasiparticle
decay rate is of the same order of magnitude as the quasiparticle energy.
We would also like to remark that according to the photoemission and tunneling
results for underdoped samples, the quasiparticle peak (with a width of order
$T$) disappears completely above $T_c$ while the gap remains at $(0,\pi)$.
Based on the slave-boson theory, the sharp electron-tunneling peak arises due
to the condensation of the holons (whose weight is proportional to $x$ at
$T=0$).  As $T$ approaches $T_c$, the fraction of the condensed holons
vanishes. If we assume that the holons are very incoherent above $T_c$, we can
conclude that the sharp electron tunneling peak should disappear above $T_c$.
This picture from the slave-boson model is completely consistent with the
observed results from photoemission experiments.

Another point to make here is about the van Hove singularity.  Samples with
small superconducting gaps ($\Delta \sim 25$meV) show a double peak structure
in the tunneling $dI/dV$ curve, and the double peak structure crosses over
into a single peak for large superconducting gaps.  At first sight, one might
guess that the double peak structure is due to the van Hove singularity.
However, after comparing the experimental lineshape with the theoretical
lineshape, we conclude that the double peak cannot arise due to the van Hove
singularity.  This is because the experimental peaks at higher bias are quite
symmetric, while the peaks from the van Hove singularity are very asymmetric
(the peak at the hole side is much stronger than the peak on the electron
side).  Based on the dispersion obtained from the fitting of the ARPES
measurements, the van Hove singularity should show up even for samples with
larger gaps ($\Delta = 50$meV). However, experimentally, one fails to see the
van Hove singularity even when the gap is as small as $20$meV.\cite{Pan} 
This seems to
suggest that quasiparticles have very short lifetime (spin-charge separation)
above the superconducting gap, and hence the van Hove singularity cannot be
observed.  This leaves us with the question of where the double-peak feature
comes from if there are no well defined quasiparticles above the
superconducting gap?  We hope that the double-peak structure may give us some
hints on how coherent quasiparticles  emerge  in the superconducting state
from the  incoherent normal state.

Finally we would like to point out that the results of this paper are obtained
from a mean-field calculation within the slave-boson theory. One naturally
questions the reliability of the mean-field result and how much of our result
remains valid after the gauge and other
fluctuations are included. One of our main
results is the explanation of the asymmetric tunneling background. It can be
traced back to the strong on-site repulsion. It is much easier to
remove an electron than to add an electron and create a doubly occupied site.
We believe this result is robust and will survive the fluctuations around the
mean-field state.  The second main result is that the weight of the coherent
quasiparticle tunneling peak is proportional to the doping $x$. After
including the fluctuations, we believe that the weight of the coherent peak
should have a similar doping dependence. This is because the coherent peak
comes from the quasiparticle which is a bound state of a spinon and a holon.
However, the detailed dependence may be of a more general form $x^{1+\alpha}$
({\it ie} the fluctuations may correct the exponent).\cite{RW}

\vspace{3mm}
We would like to acknowledge helpful discussions with J.C.Davis, D.A Ivanov,
P.A. Lee and S.H. Pan. This work is supported by NSF Grant No. DMR--97--14198
and by NSF-MRSEC Grant No. DMR--98--08941.

\end{document}